\documentclass[showpacs,10pt,aps,prd,reprint,onecolumn,notitlepage,nofootinbib,superscriptaddress]{revtex4-1}

\usepackage{hyperref}
\usepackage{amsmath}
\usepackage{mathtools}
\usepackage{bm}
\usepackage{amssymb}
\usepackage{graphicx}
\usepackage[]{subfigure}
\usepackage[usenames, dvipsnames]{color}
\usepackage{filecontents}

\hypersetup{
    bookmarks=true,         % show bookmarks bar?
    pdftoolbar=true,        % show Acrobats toolbar?
    pdfmenubar=true,        % show Acrobats menu?
    pdffitwindow=true,     %window fit to page when opened
    pdfstartview={FitH},     %fits the width of the page to the window
    pdftitle={My title},     %title
    pdfauthor={author},      %author
    pdfsubject={Subject},    %subject of the document
    pdfcreator={Creator},    %creator of the document
    pdfproducer={Producer},  %producer of the document
    pdfkeywords={keyword1} %{key2} {key3}, % list of keywords
    pdfnewwindow=true,       %links in new window
    colorlinks=true ,        %false: boxed links; true: colored links
    linkcolor=blue  ,        %color of internal links (change box color with linkbordercolor)
    citecolor=blue,      %color of links to bibliography
    filecolor=green,       %color of file links
    urlcolor=blue            %color of external links
}

\providecommand{\ed}{\mathrm{d}}

\begin{document}

\title{Congruence Convergence in pp-wave Spacetime}

\author{Mohsen Fathi}
\email{m.fathi@shargh.tpnu.ac.ir; \,\,mohsen.fathi@gmail.com}

\author{Morteza Mohseni}

 \affiliation{Department of Physics, 
Payame Noor University (PNU),
P.O. Box 19395-3697 Tehran, Iran}

\begin{abstract}
We argue that the well-known geodesic completeness property of pp-waves, can be disregarded once the geodesics are extracted as being extended along sets of Brinkmann coordinates. This issue is investigated in the more general context of congruence convergence and we show that the problem leads to diverse issues for non-geodesic congruences. The discussion is mostly based on the null congruence expansion and a generalized Raychaudhuri equation is also provided.

\bigskip

\noindent{\textit{keywords}}:  Gravitational waves, pp-waves, Congruence expansion, Generalized Raychaudhuri equation, Convergence

\end{abstract}

\maketitle

\section{Introduction}\label{sec:introduction}
Two years before Einstein proposed the general theory of relativity, he had an interesting talk with Max Born, dealing with the propagation of infinitely small perturbed fields on the base spacetime, which as Einstein argued, would travel with the speed of light \cite{Einstein2013a}. Such propagation of the \textit{gravitational action} across spacetime, was then designated as the \textit{gravitational waves}, in the sense that they would resemble some small ripples or "water waves" across the ocean \cite{MTW}. This phenomenon was originally constructed as a theoretical feature of general relativity and since then, the existence or the possibility of detection of gravitational waves have been under serious debates. This was until the recent LIGO observation of gravitational waves generated by merging black holes \cite{LIGO2016}, confirming that Einstein was indeed right. Although Einstein himself had insisted on the complexity of the field equations and hardness of finding exact solutions for propagating perturbations (specially those strong ones), nevertheless, so mush effort spurred by scientists to characterize such perturbations. As a famous approach, one recalls the linearized theory of gravitation in which the perturbations on spacetime are characterized by means of the weak-field limit of general relativity. Through this approach, the propagation equations are those which consequently provide the behavior of the wave packages in the transverse-traceless (TT) gauge. This method also gives information about the wave polarization\footnote{Most of the textbooks on general relativity provide extensive information on linearized gravity and the TT gauge. For example see relevant chapters in Refs. \cite{MTW, Ryder2009, Schutz2009}.}. However, for null fields which traverse the spacetime with the speed of light, relevant discussions were first made by Brinkmann \cite{Brinkmann1925} and Baldwin and Jeffery \cite{Baldwin1926}, who designated such fields as plane-fronted parallel waves (rays). These waves were subsequently named after as the \textit{pp-waves} by Ehlers and Kundt \cite{Witten1962}. The consequent pp-wave spacetimes are known as those exact solutions to general relativity which admit a covariant constant null vector field \cite{Stephani2003}. Essentially, one can regard the pp-waves as electromagnetic analogues of gravitation. In other words, we consider such plane waves as gravitational effects of the electromagnetic fields \cite{Holten1997}. However, as ripples on the spacetime, pp-waves are indeed gravitational plane wave solutions to general relativity\footnote{For more detailed reviews on pp-waves, in addition to those mentioned above, see Refs. \cite{Hall2004, Flores2006}.}. Ever since the advent of these exact solutions, there have been a great deal of research devoted to their characteristics, motivated by the fact that gravitational waves on their own, make an interesting topic of research both from theoretical and experimental viewpoints. These solutions have also tested in generalized and modified theories of gravity \cite{Gurses1978}\footnote{For reviews see also Refs. \cite{Baykal2016, Baykal2017}.}. They have also investigated in a 2-dimensional realm to discuss the Birkhoff's theorem \cite{Schmidt1998}. From the conformal invariance perspective, which Brinkmann discussion was based upon, researchers have also found functional form of the conformal factor \cite{Gejji2001} and also conformal symmetric classes for pp-waves \cite{Keane2004}. Even in the context of string theories and super-gravity, super-symmetric pp-waves were also discussed in a massive Penrose limit. This study specially showed that half of the super-symmetries could be recovered in the pp-wave solutions \cite{Singh2004}. Moreover, the pp-waves were generalized into the non-vacuum case to obtain non-plane forms \cite{Steele2010}. Plane wave solutions have also been put in the strong-field gravity and arbitrary polarization states were constructed \cite{Cropp2010}. The vacuum pp-wave spacetimes, were finally given a complete invariant classification, based on Cartan invariants. The authors of Ref. \cite{Milson2013} found that it is the alternation of the classifying invariants that may help us get rid of the problem of vanishing scalar curvatures and provides the possibility of doing reasonable classifications. Another interesting application of pp-wave spacetimes, is in the context of gravity plus matter. Recently, and in the Weyl-Lewis-Papapetrou structure, these plane waves have proven to be the solutions of self-gravitating Weyl fermions \cite{Cianci2015}.

All that told about the community's interest on pp-wave spacetimes, in this paper however, we concern with one of the  most basic geometrical properties of the pp-waves. An important re-known issue is the geodesic completeness of vacuum pp-waves \cite{MTW}. This means that these plane waves never experience a focusing during their propagation life-time. In this paper, we make some alternative geometrical constructions than those which are extant in the literature, to argue that this property of pp-waves vitally depends on the choice of the congruences. In other words, we can not generalize such completeness to all kinds of plane waves. This is mostly done by choosing a set of reciprocal null congruences in the spacetime manifold to specify the manifold foliations. Our intention is to put the mentioned property as a subsidiary consequence of congruence divergence/convergence behaviors and scrutinize them by means of investigating the null congruence generators of pp-waves and the transverse evolution (congruence expansion) of the propagating fields on the base manifold. Through this approach, we find that geodesic congruences have to fulfill some certain requirements to become considerable as physical null congruences. We show that because of their constant (positive or negative) expansions, even these kinds of geodesics can not be regarded as being complete. We also define a set of non-geodesic congruences and argue that this kind is capable of making considerable convergence by letting the congruence to  pass a certain boundary. These features are proven to be consequences of writing the congruence generators in terms of different sets of Brinkmann coordinates. 

The paper is organized as follows: In Sec. \ref{sec:preliminaries} we bring important mathematical notions which most of them are used extensively within the text. In Sec. \ref{sec:geodesics} we construct different sets of geodesics and discuss their transverse properties and their implications to geodesic convergence. In Sec. \ref{sec:non-geodesics} we generalize our approach to non-geodesic pp-waves and find a constraint under which the expansions can change sign. This constraint will specifically utilized in Sec. \ref{sec:surface} to show that the whole spacetime 3-dimensional foliation is itself foliation-dependent. We argue that this leads to a foliation-dependent extrinsic curvature. We summarize in Sec. \ref{sec:conclusion}. Throughout this paper we use Greek indices $\alpha, \beta, \gamma ,\dots$ to vary within $0, 1, 2, 3$, and Latin ones as $a, b, c, \dots = 1, 2 ,3$. We also use bold mathematical terms as (co)vectors.

\section{Some Mathematical Preliminaries}\label{sec:preliminaries}

In this section, we consider a congruence of gravitational null curves in the pp-wave spacetime, constructed by the tangential vector $\ell ^{\alpha }$. However, since the 4-dimensional spacetime manifold with the metric $g_{\alpha \beta }$, is characterized by the 2-dimensional light-cone, one auxiliary null vector is also necessary to define the orientation of the transverse sub-space. This null vector is notated by $n^{\alpha }$ which is commonly known as the ingoing congruence \footnote{ But as we will see in the following, it is not always the case.}. It should be noted that in this section, we consider cases in which $\bm n\cdot \bm\ell \equiv n_{\alpha } \ell ^{\alpha }$ is not necessarily constant (the case of non-normalized congruences). Also we take cases where $\bm\ell$ is not parallelly transported along the congruence (non-geodesic congruences). Hence, the two specific cases of geodesic (whether normalized or non-normalized) and non-geodesic non-normalized congruences are addressed. \\

As it was mentioned before, in this work we are interested in the transverse behavior of the congruences. First of all, let us define the following notations for different kinds of acceleration terms:
\begin{eqnarray}\label{one}
a^{\sigma }&\equiv& \dot{\ell }^{\sigma }=\ell ^{\sigma }{}_{;\rho }\ell ^{\rho },\nonumber\\
a_n^{\sigma }&=&\ell ^{\sigma }{}_{;\rho }n^{\rho },\nonumber
\\
b^{\sigma }&\equiv& \dot{n}^{\sigma }=n^{\sigma }{}_{;\rho }\ell ^{\rho },\nonumber
\\
b_n^{\sigma }&=&n^{\sigma }{}_{;\rho }n^{\rho }.
\end{eqnarray}\label{two}
Note that, considering $\tau$ as the parameterization of the congruence, there are some conventional operators as used above, which are defined in terms of their own specific notation in the following way:
\begin{eqnarray}
\cdot &\doteq&   \frac{D}{\text{d$\tau $}}~~~~~~~~,\doteq\partial\nonumber\\
\prime &\doteq&\frac{\partial }{\partial \tau }~~~~~~~~;\doteq\nabla. 
\end{eqnarray}
One other quantity, is the transverse rate of change of the congruences measured by a comoving observer; namely the expansion. Firstly, we define
\begin{equation}\label{three}
B^{\sigma }{}_{\rho } \dot{=} ~\ell ^{\sigma }{}_{;\rho },
\end{equation}
and accordingly the expansion is defined as
\begin{equation}\label{four}
\Theta =\bar{B}^{\rho }{}_{\rho }=h{_\alpha }^{\rho }h{_\sigma }^{\alpha }\ell ^{\sigma }{}_{;\rho }=h{_\sigma }^{\rho }\ell ^{\sigma }{}_{;\rho },
\end{equation}
in which
\begin{equation}\label{five}
h{_\sigma }^{\rho }=\delta_{\sigma }^{\rho }-(\bm n\cdot \bm\ell)^{-1}\left({ \ell _{\sigma }n^{\rho }+n_{\sigma } \ell ^{\rho }}\right),
\end{equation}
is the transverse projector of the tangent space (the vector field) onto the transverse sub-space of the manifold (the 2-dimensional cross-section of the congruence), which satisfies:
\begin{subequations}\label{six}
\begin{align}
\ell ^{\sigma } h{_\sigma }^{\rho }=n^{\sigma } h{_\sigma }^{\rho }=0,\\
h{_\rho }^{\rho }=\delta _{\rho }^{\sigma } h{_\sigma }^{\rho }=2.
\end{align}
\end{subequations}
Using this, the expansion of the congruence generated by $\bm\ell$ is obtained as (see appendix \ref{app:derivation})
\begin{equation}\label{I}
\Theta=\ell ^{\rho }{}_{;\rho }-(\bm n\cdot\bm\ell)^{-1}({\bm n\cdot \bm a})=\ell ^{\rho }{}_{;\rho }-(\bm n\cdot\bm\ell )^{-1}\left(\dot{(\bm n\cdot\bm\ell) }-\bm\ell \cdot \bm b\right).
\end{equation}
Note that, for $\bm n\cdot\bm\ell=\mathrm{const.}$ (normalized congruences), we have $\bm n\cdot \bm a=-\bm\ell\cdot\bm b$. One can also calculate the expansion of the congruence generated by $\bm n$, as
\begin{equation}\label{II}
\Theta _n= h{_\sigma }^{\rho }n^{\sigma }{}_{;\rho }
= n^{\rho }{}_{;\rho }-(\bm n\cdot \bm\ell)^{-1}(\bm\ell \cdot\bm{b}_n).
\end{equation}
For normalized congruences, it is $\bm\ell\cdot\bm b_n=-\bm n\cdot\bm a_n$.  One imporatant mathematical tool, for probing the singularity theorems, is the Raychaudhuri equation, which gives the evolution of the transverse rate of change of the congruence generated by $\bm\ell$, i.e. the $\dot\Theta$ quantity. Following the method of Ref. \cite{Thompson2017}, we derive the generalized Raychaudhuri equation for non-geodesic, non-normalized, null congruences (see Appendix \ref{app:derivation} for details).  The result is
\begin{eqnarray}\label{III}
\dot{\Theta }&=&-\frac{1}{2} \Theta ^2-\sigma _{\mu \nu }\sigma ^{\mu \nu }+\omega _{\mu \nu }\omega ^{\mu \nu }-R_{\mu \nu } \ell ^{\mu } \ell ^{\nu }+a^{\mu }{}_{;\mu }-a^{\mu }{}_{;\nu }\left(\delta _{\mu }^{\nu }-h{_\mu }^{\nu }\right)\nonumber\\
&&-({\bm\ell \cdot\bm n})^{-1}\left[2 n_{\nu }\ell ^{\nu }{}_{;\mu }a^{\mu }+(\bm a\cdot \bm a_n)+(\bm b\cdot\bm a)\right]+(\bm\ell \cdot\bm n)^{-2}\left[(\bm n\cdot\bm a)^2+(\bm n\cdot\bm a) \dot{(\bm \ell \cdot\bm n)}\right],
\end{eqnarray}
where $\Theta$ has been defined in Eq. (\ref{I}). To define $\sigma_{\mu\nu}$ and $\omega_{\mu\nu}$, respectively as the shear and vorticity tensors, let us decompose $\bar{B}_{\mu \nu }=g_{\mu \lambda } \bar{B}^{\lambda }{}_{\nu }$ into its symmetric and anti-symmetric parts, as below \cite{Poisson2009}:
\begin{equation}\label{eight}
\bar{B}_{\mu \nu }=\bar{B}_{(\mu \nu) }+\bar{B}_{[\mu \nu ]}=\theta _{\mu \nu }+\omega _{\mu \nu },
\end{equation}
and the symmetric part into trace and traceless segments,
\begin{equation}\label{nine}
\theta _{\mu \nu }=\frac{1}{2}{\Theta  h_{\mu\nu} }+\sigma _{\mu \nu },
\end{equation}
with $h_{\mu \nu }=g_{\mu \lambda } h^{\lambda }{}_{\nu }$, giving $\theta ^{\mu }{}_{\mu }=\Theta$ since $\sigma ^{\mu }{}_{\mu }=0$. Having these, the mentioned tensors are defines as
\begin{subequations}\label{IV}
\begin{align}
\sigma _{\mu \nu }= h{_\mu }^{\sigma } h{_\nu }^{\lambda } \ell _{(\sigma ;\lambda) }-\frac{1}{2}{\Theta  h_{\mu \nu }},\\
\omega _{\mu \nu }=h{_\mu }^{\sigma }h{_\nu }^{\lambda }\ell _{[\sigma ;\lambda ]}.
\end{align}
\end{subequations}
The Raychaudhuri equation in Eq. (\ref{III}), is rather general. But the version which is extant in the literature is that of normalized geodesic congruences, either for affinely or non-affinely parameterzied ones. For non-affinely parameterized geodesics, we have $\bm n\cdot\bm\ell=\mathrm{const.}$, and  $\bm a=\kappa\bm \ell$, giving an alternative defintion for the surface gravity $\kappa$ \cite{Faraoni2015}. Using Eq. (\ref{I}), the expansion becomes 
\begin{equation}\label{V}
\Theta =\ell ^{\rho }{}_{;\rho }-\kappa,
\end{equation}
the evolution of which is governed by the famous equation \cite{Faraoni2015, Kar2007}
\begin{equation}\label{VI}
\dot\Theta = -\frac{1}{2}\Theta^2 - \sigma_{\mu\nu}\sigma^{\mu\nu}+\omega_{\mu\nu}\omega^{\mu\nu}-R_{\mu\nu}\ell^\mu\ell^\nu+\kappa \Theta.
\end{equation}
 Note that, the generalized version in Eq. (\ref{III}) is reduced to Eq. (\ref{VI}) for normalized non-affinely parameterized geodesic congruences (see appendix \ref{app:derivation}).  For affinely parameterized geodesics, one puts $\kappa=0$ and gets to the most common version of the Raychaudhuri equation for geodesic null congruences with $\Theta =\ell ^{\rho }{}_{;\rho }$. One can see that, the only item that changes $\Theta$ and $\Theta_n$ form those their well-known forms, is the existence of acceleration terms, not the possibility of $\bm n\cdot\bm\ell\neq\mathrm{const.}$ We should point out that the applicability of the above expansions, goes beyond the determination of the cross-sectional shape of the congruences. They can even be used to characterize regions on which the congruences switch their behavior. To give more details, we mention the trapped surfaces, on which the null congruence is trapped. Indeed, a trapped surface refers to the light cone itself. More precisely, at some particular point along the reference congruence, the expansion vanishes and  basically the light cone closes up. Now considering all other congruences and repeating the procedure, this will identify a locus of spacetime points at which the light cone is \textit{marginally trapped}  and will therefore define a spacetime surface that is referred to as a \textit{marginally trapped surface}. We can replace "marginally trapped" with "trapped" when the outgoing expansion is negative instead of vanishing. To bring more information, a surface on which $\Theta\Theta_n>0$, can be trapped or anti-trapped, whereas a 2-surface satisfying $\Theta\Theta_n<0$ is called un-trapped \cite{Faraoni2015}. Whether these surfaces are null, space-like, or time-like  depends on the particular locus of points (through the property of the tangent or normal to the 2-dimensional surface) \footnote{The 3-dimensional closure of all these trapped surfaces is called the apparent horizon.}.

\subsection{Geodesic Raychaudhuri Equation in pp-wave Spacetime}\label{sebsec:geodesicRaychaudhuri}

\begin{figure}[htp]
\begin{center}
\includegraphics[width=5cm]{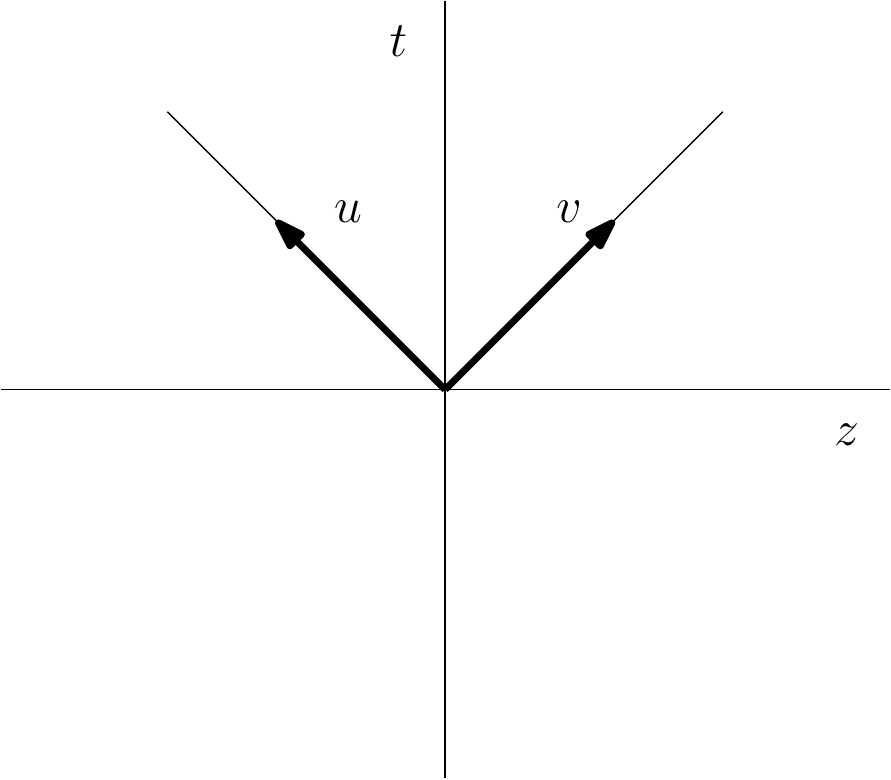}
\caption{\label{fig:NullCone} The outgoing/ingoing future-directed extensions in the Brinkmann coordinates.}
\end{center}
\end{figure}

The pp-wave spacetime is a plane wave solution to Einstein field equations, admitting the metric, which in the Brinkmannn coordinates $(u,v,x,y)$ reads as \cite{Steele2010}
\begin{equation}\label{ten}
\ed s^2 = 2 H(u,x,y) \ed u^2 + 2 \ed u \ed v + \ed x^2 + \ed y^2,
\end{equation}
in which the $\bm v$ coordinate is itself a null vector field, stretched along the outgoing geodesics. The coordinate $\bm u$  can be related to null, time-like or space-like vector fields. In this work however, $\bm u$ and $\bm v$ are supposed to constitute the cone sides of Fig.~\ref{fig:NullCone}, which here are considered as coordinates in the system $x^\alpha(\tau) = \Big(u(\tau),v(\tau),x(\tau),y(\tau)\Big)$, and are defined as
\begin{subequations}\label{eleven}
\begin{align}
u = t - z,\\
v = t + z.
\end{align}
\end{subequations}
Let $\lambda_\alpha=u_{,\alpha}=(1,0,0,0)$ to be a covector field in the cotangent bundle. This field is of course along the $\bm v$ coordinate. The $\bm\ell$ congruence is spanned by $x_{,\alpha}$ and $y_{,\alpha}$. Hence, the null vectors $\bm\ell$ and $\bm n$ are those which comprise the null cone. The gravitational potential $H(u,x,y)$ can be written as \cite{Steele2010}
\begin{equation}\label{twelve}
H=-\frac{1}{2}g^{\alpha \beta }v_{,\alpha }v_{,\beta }
\end{equation}
and therefore it can somehow be regarded as a null Hamiltonian of the form $\mathcal{H}=-\frac{1}{2} g^{\alpha \beta } k_{\alpha } k_{\beta }$ where $k_\alpha\doteq v_{,\alpha}$ is the covector field associated with the congruence generated by $\bm\ell$ and is indeed a member of the cotangent fiber bundle of the spacetime manifold. The pp-wave spacetime also implies $g^{\alpha \beta }H_{,\alpha }\lambda _{,\beta }=0$ so that $H$ is independent of $\bm v$. Furthermore, the Ricci tensor is obtained as $R_{\alpha \beta }=-g^{\rho \sigma }H_{;\rho \sigma }\lambda _{\alpha }\lambda _{\beta }=-\left(H_{,{xx}}+H_{,{yy}}\right)\lambda _{\alpha }\lambda _{\beta }$ and therefore $R_{\mu \nu } \ell ^{\mu } \ell ^{\nu }=-g^{\rho \sigma }H_{;\rho \sigma }(\bm\lambda\cdot\bm\ell )^2$ for which the positivity or negativity of this term depends explicitly on the sign of $H{^{;\rho}}_{;\rho}$. Note that, $R_{\mu\nu}\ell^\mu\ell^\nu$ is the crucial term to discuss the strong energy condition (SEC) in general relativity. For $R_{\mu \nu } \ell ^{\mu } \ell ^{\nu }\geq 0$, the SEC is satisfied and gravity is an attractive force.  In this work, and in the context of general relativity, we respect the SEC and therefore, we consider $H^{;\rho }{}_{;\rho } \leq 0$. Now for affinely parameterized geodesic congruences, the Raychaudhuri equation in Eq. (\ref{VI}) can be recast by calculating $\dot{\Theta }=\frac{D}{\ed\tau}\left(\ell ^{\mu }{}_{;\mu }\right)$ as below:
\begin{equation}\label{VIII}
\dot{\Theta }=-\ell ^{\mu }{}_{;\nu }\ell ^{\nu }{}_{;\mu }+g^{\rho \sigma }H_{;\rho \sigma }(\bm\lambda \cdot\bm\ell )^2.
\end{equation}

\section{Geodesic Congruences in pp-wave Spacetime}\label{sec:geodesics}

The geodesic equation $\bm a = \bm 0$, is equivalent to $\left(\ell ^{\alpha }\right)'+ \Gamma ^{\alpha }{}_{\rho \lambda } \ell ^{\rho } \ell ^{\lambda }=\bm 0$ with $\ell^\alpha = x^{\prime\alpha}$. In pp-wave spacetime, the geodesic equations are
\begin{subequations}\label{IX}
\begin{align}
u''=0,\label{IXa}\\
v''+u'\left(u'H_{,u}+2x'H_{,x}+2y'H_{,y}\right)=0,\label{IXb}\\
x''-u^{\text{$\prime $2}}H_{,x}=0,\label{IXc}\\
y''-u^{\text{$\prime $2}}H_{,y}=0\label{IXd}.
\end{align}
\end{subequations}
The null condition has also to be satisfied by the congruences, giving the constraint
\begin{equation}\label{X}
2 H u'+2 u' v'+x^{\text{$\prime $2}} + y^{\prime 2} = 0.
\end{equation}

\subsection{ $\mathbf{z}$-directed Geodesics}\label{subsec:z-directed}

The geodesic equations and the null condition imply $u'=E=\text{const}.$ and $v'=-E^2H_{,u}$ whereas the third and the fourth geodesic equations require $H_{,x}=H_{,y}=0$. Hence, to have a $z$-directed geodesic congruence, we need $H\equiv H(u)$. We get accordingly,
\begin{equation}\label{XI}
\ell ^{\alpha }=(E,-E H(u),0,0).
\end{equation}
The above vector generates a parallelly transported outgoing geodesic congruence but it does not coincide to the $\bm v$ coordinate. Also for $\bm\ell$ to be future directed, one needs $E<0$ and $H(u)>1$ for every $u$. We now look for a relevant auxiliary null vector $\bm n$. The condition $\bm n\cdot \bm n = 0$, provides
\begin{equation}\label{XII}
n^{\alpha }=\left(n^0,- n^0 H(u)  ,0,0\right),
\end{equation}
in which $n^0$ is not necessarily constant. Moreover, since $H(u)>1$, there is no way of having a future directed ingoing $\bm n$. If such congruence is supposed to be ingoing, then we should let $n^0>0$. This surely gives a past directed congruence toward $-z$ and $-t$\footnote{We should here note that when we say the vector is stretched along $-z$, it does not mean that its $z$ component is negative; it means that its arrow points toward the direction along which, $z$ decreases. We can also designate the origin, as the point in the present ($t = 0$).}. These two vectors also give $\bm n\cdot\bm\ell = 0$ which is undesirable, because this way, these vectors can not describe a transverse subspace. The reason is because from Eqs. (\ref{XI}) and (\ref{XII}) we see that $\bm n = \frac{n^0}{E} \bm\ell$. Accordingly, for $\frac{n^0}{E}>0$, the congruence generated by $\bm n$ is future directed but it also completely lies on the same direction of $\bm\ell$. Similarly, for $\frac{n^0}{E}<0$, it is past directed. So these vectors are along each other, either in opposite or the same directions. Hence, a future-$z$-directed outgoing geodesic in pp-wave spacetime, corresponds to a past-$z$-directed ingoing congruence, with no reasonable normalization.

\subsection{$\mathbf{zx}$-directed Geodesics}\label{subsec:zx}

In this case, the fourth equation of Eqs. (\ref{IX}), implies $H\equiv H(u,x)$. But from Eq. (\ref{IXc}) and knowing $u^\prime=E$, we have
\begin{eqnarray}\label{fourtheen}
x' x''=E^2x'H_{,x}~~~~&\Longrightarrow&~~~~ x' x''=E^2 H'\nonumber\\
&\xRightarrow[\text{using}~x''=E^2H_{,x} ]{ }&~~~~x'\left(E^2H_{,x}\right)=E^2\left(x'H_{,x}+b^2H_{,u}\right)\nonumber\\
E^2H_{,u}=0~~~~&\Longrightarrow&~~~~ H_{,u}=0.
\end{eqnarray}
This means that if the geodesic equations are to be respected, then we must expect  $H\equiv H(x)$. Therefore from Eqs. (\ref{IXa}), (\ref{IXc}) and (\ref{X}), we obtain 
\begin{equation}\label{XIII}
\ell ^{\alpha }=\left(E,-2 E H(x),\pm E \sqrt{2 H(x)},0\right),
\end{equation}
as the generator of a $zx$-directed outgoing geodesic congruence. Here we impose the condition $H(x)>1/2$ for the congruence to be future directed (i.e. toward $+t$)\footnote{We can be assured that since $H$ is inside a square root, it must be positive. Also for $\bm\ell$ to point toward $+\bm v$, $H$ and $E$ must have opposite signs; hence, $E<0$.} and to propagate toward $+z$. However, the distinction occurs once we take into account, the plus and minus signs of the third component of $\bm\ell$. The plus sign gives a congruence moving toward $-x$ direction whereas the minus sign, gives one along  $+x$. So, for such congruences, instead of the $z$ direction, we are able to designate $\pm x$ as directions for outgoing/ingoing congruences and build the generators upon them. Here, since we are interested in an outgoing $\bm\ell$, we adopt the negative sign. Now, applying the conditions $\bm n\cdot\bm n=0$ and $\bm n\cdot\bm\ell=-1$, we obtain a relevant auxiliary congruence for $\bm\ell$. These conditions result in the following two copies of $\bm n$:
\begin{subequations}\label{XIV}
\begin{align}
n_1^\alpha=\left(n^0,\frac{1}{E}\left[{-2 E n^0 H(x)+2 \sqrt{E n^0 H(x)}-1}{}\right],\sqrt{2} \left[\sqrt{\frac{n^0}{E}}-n^0 \sqrt{H(x)}\right],0\right),\label{XIVa}\\
n_2^\alpha=\left(n^0,-\frac{1}{E}\left[{2 E n^0 H(x)+2 \sqrt{E n^0 H(x)}+1}{}\right],-\sqrt{2} \left[\sqrt{\frac{n^0}{E}}+n^0 \sqrt{H(x)}\right],0\right).\label{XIVb}
\end{align}
\end{subequations}
In both cases, one can see that $E$ and $n^0$ have to be of the same sign. Therefore we infer $n^0<0$ (whether being a constant or not). However, the above constraints on the values, stretch $\bm n_1$ along $+z$, $+t$ and $+x$. Therefore it is not ingoing. This distinction goes to $\bm n_2$ (see appendix \ref{app:congruences}). The mentioned constraints stretch it along $+z$, $+t$ and $-x$. In this regard, it makes sense to talk about this particular congruence and its transverse behavior. Note that, the above congruences, satisfy $n^0 E <2$. We can confirm that the geodesic Raychaudhuri equation (\ref{VIII}) is satisfied for the $\bm\ell$ in Eq. (\ref{XIII}). Now let us calculate the expansions. Using Eqs. (\ref{V}) with $\kappa=0$ and Eq.~(\ref{II}), together with Eqs. (\ref{XIII}) and (\ref{XIVb}) we have
\begin{subequations}\label{twenty-three}
\begin{align}
\Theta = -\frac{E H_{,x}}{\sqrt{2 H(x)}},\label{twenty-threea}\\
\Theta_{n_2} = \frac{n^0 H_{,x}}{\sqrt{2 H(x)}},\label{twenty-threeb}\\
\dot\Theta = -\frac{E^2\left(H_{,x}\right){}^2}{2 H(x)}+E^2H_{,{xx}}.\label{twenty-threec}
\end{align}
\end{subequations}
For $H_{,x}>0$, then $\Theta>0$ and $\Theta_n<0$, and if  $H_{,x}<0$, then $\Theta<0$ and $\Theta_n>0$. Both cases give correspondence to an un-trapped region where the gravitational waves are being propagated. So, the congruences can travel freely, inwardly and outwardly. The condition $\Theta\Theta_n<0$ here, however points to the fact that the divergence of one congruence, gives rise to the convergence of its reciprocal counterpart. This may be in contrast with the well-known extant agreement that geodesic pp-waves avoid any geodesic incompleteness. In fact, this latter can only happen once the outgoing null congruence is set to be completely the $\bm v$ coordinate. On the other hand, one important thing that we should bear in mind is the implications of the $\dot\Theta$ value. In our current case of study, the SEC provides $H_{,xx}<0$. This surely makes the $\dot\Theta$ in Eq. (\ref{twenty-threec}) to be of a negative value. This is an inference of attractive gravity in the context of general relativity. Hence, although in the region filled by the geodesic gravitational wave congruences we do not encounter any incompleteness, nevertheless, the outgoing congruence always experiences a convergent force (either for those expanding or those contracting). Therefore, focusing of the outgoing geodesic congruences under study, would be inevitable along the $x$ direction within a finite time.

\section{Non-geodesic Normalized Congruences}\label{sec:non-geodesics}

A pair of normalized non-geodesic congruences in the $z, x$ directions, can be let to be the same as those in Eqs. (\ref{XIII}) and (\ref{XIV}), but by taking $H\equiv H(u,x)$. All the previous conditions we had for having future directed outgoing/ingoing congruences along $\pm x$, do hold here. This way, we let
\begin{equation}\label{XIX}
\ell^\alpha = \left(E, -2 E H(u,x),  -E \sqrt{2 H(u,x)}, 0\right),
\end{equation} 
\begin{equation}\label{XX}
n^\alpha=\left(n^0,-\frac{1}{E}\left[{2 E n^0 H(u,x)+2 \sqrt{E n^0 H(u,x)}+1}{}\right],-\sqrt{2} \left[\sqrt{\frac{n^0}{E}}+n^0 \sqrt{H(u,x)}\right],0\right),
\end{equation}
satisfying $\bm n\cdot \bm\ell = -1$, and giving the acceleration terms
\begin{subequations}\label{XXI}
\begin{align}
a^\alpha = \left(0, -E^2 H_{,u}, -\frac{E^2 H_{,u}}{\sqrt{2 H(u,x)}}, 0\right),\label{XXIa}\\
a_n^\alpha = \left(
0, E\left[\sqrt{\frac{2 n^0}{E}} H_{,x}-n^0 H_{,u}\right],
 \frac{E\left[2\sqrt{\frac{ n^0}{E}} H_{,x}-\sqrt{2}n^0 H_{,u}\right]}{2\sqrt{H(u,x)}}, 0
\right),\label{XXIb}\\
b^\alpha = \left(
0, -\frac{E\left[n^0 H_{,u}\left(\sqrt{E n^0 H(u,x)}+1\right)\right]}{\sqrt{E n^0 H(u,x)}},
 -\frac{E n^0 H_{,u}}{\sqrt{2 H(u,x)}},
  0
\right),\label{XXIc}\\
b_n^\alpha = 
\left(0, \frac{{n^0} \left[\sqrt{2} H_{,x} \left({n^0} \sqrt{H(u,x)}+\sqrt{\frac{{n^0}}{{E}}}\right)-{n^0} H_{,u} \left(\sqrt{ {E} {n^0} H(u,x)}+1\right)\right]}{\sqrt{{E} {n^0} H(u,x)}}, 
\frac{{n^0} \left[\sqrt{2} {n^0} H_{,u}-2 \sqrt{\frac{ {n^0}}{ {E}}} H_{,x}\right]}{2 \sqrt{H(u,x)}}, 0 \right).\label{XXId}
\end{align}
\end{subequations}
One can confirm that the above quantities, together with the congruences in Eqs. (\ref{XIX}) and (\ref{XX}), satisfy the generalized Raychaudhuri equation in Eq. (\ref{III}), with $\dot{(\bm n\cdot\bm\ell)}=0$. Now that all these acceleration terms are non-zero, one should apply the generalized equations for expansions, i.e. Eqs. (\ref{I}) and (\ref{II}), to obtain the transverse rate of change of the cross-sectional area. According to Eqs. (\ref{XIX}) and (\ref{XX}), we obtain
\begin{equation}\label{XXII}
\Theta = -\frac{E \left(\sqrt{2} H_{,x}  - 2\sqrt{E n^0} H_{,u}\right)}{2 \sqrt{H(u,x)}},
\end{equation}
\begin{equation}\label{XXIII}
\Theta_n = -\frac{n^0 \left(2 E n^0 H_{,u}-\sqrt{2 E n^0} H_{,x}\right)}{2 \sqrt{E n^0 H(u,x)}}.
\end{equation}
We can inspect the above relations for some conditions for which $\Theta$ and/or $\Theta_n$ do vanish. It turns out that both of them vanish identically for
\begin{equation}\label{XXIV}
H_{,u} = \frac{H_{,x}}{\sqrt{2 E n^0}}.
\end{equation}
As both cases happen simultaneously, non of the known trapped/anti-trapped surfaces can form here, because at the same time, no congruence can emerge or enter the region. In other words, in the spacial cases of Eq. (\ref{XXIV}), there is a region which is completely isolated from the whole manifold\footnote{One also finds vanishing Lie derivatives $\mathcal{L}_{\bm n}\Theta$ and $\mathcal{L}_{\bm\ell}\Theta_n$ in accordance to this special case. This points to a serious difference  from the definition of an apparent horizon, since non-vanishing Lie derivatives give rise to the location of a trapped/anti-trapped observer (inner or outer) \cite{Faraoni2015} who can not exist for both cases at the same time.}. Another implication, inferred from the value in Eq. (\ref{XXIV}) is that by any slight deviation from this value, one can expect convergence or divergence of the outgoing/ingoing pp-waves. In fact, we can categorize the following cases:

\begin{itemize}
\item{For $H_{,u}<\frac{H_{,x}}{\sqrt{2 E n^0}}$ we have $\Theta>0$ and $\Theta_n<0$,}

\item{For $H_{,u}>\frac{H_{,x}}{\sqrt{2 E n^0}}$ we have $\Theta<0$ and $\Theta_n>0$.}       
\end{itemize}
Both cases do correspond to un-trapped regions, and therefore, the case of vanishing expansions means a shift in the behavior of the congruences. In this regard, the expansion-less condition, itself, conceals converging behaviors for both congruences. The condition (\ref{XXIV}) provides the simple solution
\begin{equation}\label{XXVI}
H(u,x)=\frac{u}{\sqrt{2 E n^0}}+x.
\end{equation}
This solution corresponds to an expansion-less pair of reciprocal congruences $(\bm n, \bm\ell)$. We can look for the characteristics of a manifold (wave package), with this particular metric potential. To elaborate this, in the next section, we talk about the characteristics of a hypersurface on which the congruences are induced. Through this method, the expansion-less hypersurface is also discussed.

\section{Characteristics of the Expansion-less Hypersurface}\label{sec:surface}

The region in which $\Theta=\Theta_n=0$ is satisfied, strictly corresponds to a definite condition on differentials of $H(u,x)$; as pointed out in Eq. (\ref{XXIV}). Here, this region is assumed to be a hypersurface $\Sigma$. In other words, we consider a hypersurface on which the above mentioned condition is satisfied for $H(u,x)$.  In this section, we mostly deal with the 3-dimensional foliation of spacetime, in accordance with the non-geodesic outgoing/ingoing congruences introduced in the previous section. Mathematically, $\Sigma$ is indeed a hypersurface with a specific metric and extrinsic curvature which we will obtain. Furthermore, we inspect the characteristics of this surface, according to the mentioned expansion-less condition.

\subsection{General Discussion}\label{subsec:surface-general}

\begin{figure}[htp]
\begin{center}
\includegraphics[width=3cm]{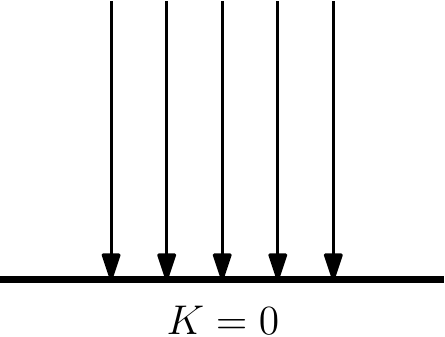}~(a)
\hfil
\includegraphics[width=3cm]{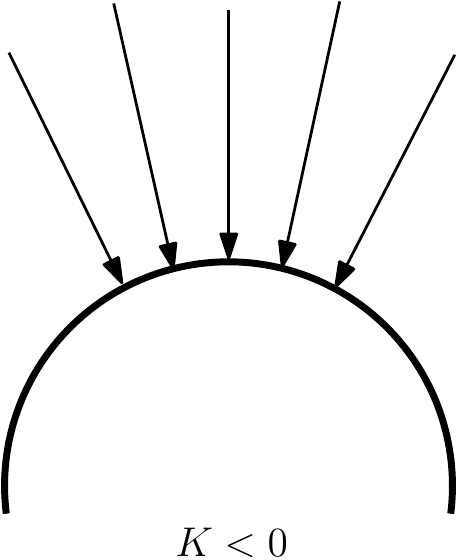}~(b)
\hfil
\includegraphics[width=3cm]{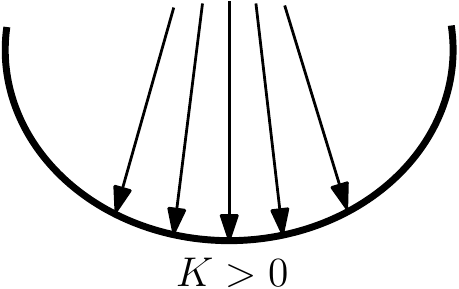}~(c)
\caption{\label{fig:extrinsic} The impacts of the different values for $K$ on the behavior of the incident congruence.}
\end{center}
\end{figure}

First of all, we note that this hypersurface is characterized by the following tangential vectors to identify the displacements on $\Sigma$:

\begin{equation}\label{XXVII}
\epsilon ^{\alpha }{}_a=\frac{\partial x^{\alpha }}{\partial y^a},
\end{equation}
in which  $y^a=(\tau,y,z)$ are the intrinsic coordinates on $\Sigma$. Note that $\tau$ is the local parameterization of $\Sigma$ which is as well, the parameterization of the outgoing/ingoing congruences. Also the $y^a$ coordinates are chosen in accordance with the fact that the outgoing/ingoing congruences are defined with respect to the $\pm x$ directions. The above tangential vectors can be regarded as projectors to pull-back 2-forms from the 4-dimensional spacetime manifold, to 2-forms on $\Sigma$. Therefore, these tangential vectors enable us to construct the induced metric $\gamma_{ab}$, in the following way:
\begin{equation}\label{XXVIII}
\gamma _{a b}=g_{\alpha \beta } \epsilon ^{\alpha }{}_a \epsilon ^{\beta }{}_b,
\end{equation}
which characterizes the line element on $\Sigma$ as $\ed s_{\Sigma }^2=\gamma _{a b}\ed y^a\ed y^b$. Now using Eq. (\ref{XXVII}), on gets
\begin{subequations}\label{XXX}
\begin{align}
\epsilon ^{\alpha }{}_1=x^{\prime \alpha }=\ell ^{\alpha },\label{XXXa}\\
\epsilon ^{\alpha }{}_2=(0,0,0,1),\label{XXXb}\\
\epsilon ^{\alpha }{}_3=(-1,1,0,0).\label{XXXc}
\end{align}
\end{subequations}
Applying Eq. (\ref{XXVIII}) together with Eqs. (\ref{XXX}), we obtain
\begin{equation}\label{XXXI}
\gamma_{ab} = 
\left( {\begin{array}{ccc}
   0 & 0 & E \\
   0 & 1 & 0 \\
   E & 0 & 2 \left(H(u,x)-1\right)
  \end{array} } \right),
\end{equation}
which by means of $\gamma^{ab}=(\gamma_{ab})^{-1}$, we get $\gamma{^a}_a=3$, confirming that $\Sigma$ is 3-dimensional. The fact that, despite of the null generators, $\Sigma$ is 3-dimensional and also $\bm\ell\cdot \boldsymbol\epsilon_A \neq\bm 0$ \footnote{In this discussion, $A, B=y,z$.}, makes us infer that it is the cross terms in the spacetime metric and the $x$-directed component in $\bm\ell$, which have serious impacts on the dimensionality of the hypersurface. We are also supposed to find a normal vector to the surface to deal with its curvature. However, as it is apparent from the fact that $\bm\ell$ is tangent to curves on this surface ($\Sigma$ is null), the orthogonality of the mentioned normal vector to $\boldsymbol\epsilon_1\equiv\bm\ell$ is not meaningful, because $\bm\ell$ is already self-orthogonal. Therefore, this orthogonality condition is utilized on the space-like part of $\Sigma$; i.e. those 2-surfaces $\mathcal{S}$, which are characterized by a time-like normal covector $N_\alpha$. These 2-surfaces, foliate the 3-surface $\Sigma$. Such covector, is obtained by satisfying the conditions $\bm N\cdot \boldsymbol\epsilon_A=\bm0$ and $\bm N\cdot\bm N=-1$. Applying these conditions, one can get
\begin{subequations}\label{XXXII}
\begin{align}
N_\alpha = \frac{1}{\sqrt{2 (H(u,x)-1)}}(1, 1, 0, 0),\label{XXXIIa}\\
N^\alpha = \frac{1}{\sqrt{2 (H(u,x)-1)}} (1, 1-2 H(u,x), 0, 0),\label{XXXIIb}
\end{align}
\end{subequations}
implying $H(u,x)>1$. So, we have got a 3-dimensional null hypersurface $\Sigma$, foliated by space-like 2-surfaces $\mathcal{S}$. However, in some way, we can still think of $\bm N$ as a normal covector to $\Sigma$, in the sense that at a particular segment (foliation), the extrinsic curvature of $\Sigma$ can be characterized by $\bm N$. This shows the foliation dependence of the hypersurface on which any specific conditions for pp-waves are supposed to be satisfied, which in our case, are the vanishing expansions. This foliation dependence, is also inferred from the extrinsic curvature of the hypersurface. Traditionally, the extrinsic curvature is a 3-tensor, showing the convexity/convcavity of the hypersurface. In our approach, the 3-tensor of the extrinsic curvature, is calculated by $K_{ab} = N_{\alpha;\beta} \epsilon{^\alpha}_a \epsilon{^\beta}_b$.
However, since we have confined ourselves on a foliating 2-surface, the extrinsic curvature is indeed 2-dimensional and is calculated by:
\begin{equation}\label{XXXIV}
K_{AB} = N_{A|B} = N_{\alpha;\beta} \epsilon{^\alpha}_A \epsilon{^\beta}_B,
\end{equation}
in which $N_{A|B}$ is the 2-dimensional covariant derivative on $\mathcal{S}$ and $N_A = N_\alpha \epsilon{^\alpha}_A$. The above extrinsic curvature 2-tensor has the only non-zero component
\begin{equation}\label{XXXV}
K_{zz} = -\frac{H_{,u}}{\sqrt{2 (H(u,x)-1)}}.
\end{equation}
This shows that everywhere on $\Sigma$, and according to the specific foliation offered by Eqs. (\ref{XXXb}) and (\ref{XXXc}), one can expect the convexity/concavity properties given by Eq. (\ref{XXXV}). Different foliations would therefore lead to different curvatures. To obtain the inverse tensor $K^{ A B}$, we define the induced space-like 2-metric tensor
\begin{equation}\label{XXXVI}
m_{AB}=g_{\alpha\beta} \epsilon{^\alpha}_A \epsilon{^\beta}_B,
\end{equation}
and the metric on $\mathcal{S}$ as $\ed s^2 = m_{AB}\ed y^A \ed y^B$. Applying Eq. (\ref{XXXVI}) together with Eqs. (\ref{XXXb}) and (\ref{XXXc}), yields
\begin{equation}\label{XXXVII}
m_{AB}=
\left( {\begin{array}{cc}
   1 & 0  \\
   0 & 2(H(u,x)-1)  
  \end{array} } \right),
\end{equation}
and therefore one obtains $K^{zz} = - \frac{H_{,u}}{[2 (H(u,x)-1)]^{5/2}}$ and also its trace, as below:
\begin{equation}\label{XXXVIII}
K = m^{A B} K_{A B} = N{^A}_{|A} = -\frac{H_{,u}}{2\sqrt{2} (H(u,x)-1)^{3/2}},
\end{equation}
where 
\begin{equation}\label{2D-expansion}
N{^A}_{|A} =~ ^2\Theta_N,
\end{equation}
is the 2-dimensional expansion of the congruence associated with $\bm N$\footnote{Even though $\bm N$ does not provide a geodesic congruence, however its expansion is still given by its divergence (which here is 2-dimensional). For more details on expansions of non-geodesic and also non-normalized time-like congruences, see Ref.~\cite{Fathi2016}.}. In general, $K$ identifies the scalar exterior curvature of a hypersurface, on which an incident time-like congruence is ingoing. Each of the vanishing, positive and negative values for $K$, designate the hypersurface as being respectively flat, convex or concave \cite{Poisson2009}. According to Eq. (\ref{2D-expansion}), one can observe that such properties could be related to expansion-less, diverging or converging incident congruences (see Fig. \ref{fig:extrinsic}).
For the special case of expansion-less \textit{null} congruences, discussed in the previous section, this scalar has been obtained in Eq. (\ref{XXXVIII}). However, to identify the convergence/divergence of the incident \textit{time-like} congruences on the foliating 2-surfaces, we use the relevant solution for the gravitational potential.

\subsection{Specific to the Expansion-less Solution of $H(u,x)$}\label{subsec:surface-specific}

In the last section, we obtained a condition for the gravitational potential $H(u,x)$ in Eq. (\ref{XXVI}), corresponding to an expansion-less set of ingoing/outgoing pp-wave congruences. As it was pointed out in the previous subsection, the convexity/concavity of the hypersurface generated by the null generators, in the pp-wave construction, depends indispensably on the foliating 2-surfaces. These properties also provide information about the possibility of divergence/convergence of the congruences. For the expansion-less pp-waves, and by means of Eqs. (\ref{XXVI}), (\ref{XXXVIII}) and (\ref{2D-expansion}), we calculate
\begin{equation}\label{XXXIX}
^2\Theta_N = -\frac{1}{4} \left(E n^0\right)^{-1/2}\left(x + \frac{u}{\sqrt{2 E n^0}} - 1\right)^{-3/2}.
\end{equation}
This confronts us with a basic situation, because everything now depends on the sign of the expression in the second parenthesis. Clearly, one would expect it to be positive, therefore one gets $^2\Theta_N<0$, the foliating 2-surface under consideration is concave and a convergence is expectable. The congruence defined in Eq. (\ref{XXXIIb}) intersects $\mathcal{S}$ orthogonally. So, one should be aware that this 2-dimensional expansion is indeed associated with the rate of changes in the area of the contours on $\mathcal{S}$ which are generated by the intersection of the incident congruence and the surface. The above negative value ensures us that once again, we encounter a converging congruence (although non-null) in pp-wave spacetime.

\section{Summary and Conclusion}\label{sec:conclusion}

The importance of congruence focusing became more highlighted, after proposing the Penrose-Hawking singularity theorems \cite{Penrose1965, Hawking1965, Hawking1966}. The theorems were put into rigorous usage in the theory of gravitational collapse, black holes and most importantly the cosmic censorship scenario. Dealing with these notions, one can also enjoy some fundamental mathematical tools as specifiers of kinematical evolution of infalling congruences (either null or time-like) in gravitational fields. One such an outstanding tool, is the Raychaudhuri equation which its effects on the energy conditions and focusing theorems have also pointed out by Penrose himself \cite{Penrose2002}. In this paper, one significant outcoming issue from above mathematical tools, namely the congruence convergence, was discussed. This discussion devoted to those null congruences which specify a region affected by gravitational waves, i.e. the pp-wave spacetime. The importance of this investigation was that any convergence could lead to a region in which no wave could enter. As we mentioned early in this study, geodesic completeness of pp-waves is a rather well agreed-upon concept. We showed however that if the null congruences are mainly based on some combinations of Brinkmann coordinates, we can expect an inevitable convergence and/or a focusing within finite time. We first presented some mathematical tools, including generalized expansion relations and a generalized Raychaudhuri equation for non-geodesic non-normalized congruences. We argued that, in our formalism, it is only under certain conditions that reasonable physical causality can be respected. Moreover, we showed that even in such cases, geodesic focusing is expectable. We extended our discussion to non-geodesic pp-waves and found out that we can construct a region (or a hypersurface) on which the outgoing/ingoing congruences experience a shift in their propagation manner. This hypersurafce was inspected according to its extrinsic curvature and we argued that in pp-wave spacetime, these surfaces are foliation dependent and even time-like incident congruences experience remarkable convergence. In conclusion, we note that if the gravitational waves are regarded as null perturbed fields, then one may encounter some regions in the spacetime manifold, where the fields exhibit reasonably particular behaviors. Here in our discussion, we encountered regions of expansion-less pp-waves. These results may help us improving our insights into the notion of gravitational waves, as ripples on the spacetime ocean. 

\appendix

\section{Derivation of the Generalized Raychaudhuri Equation for Null Congruences}\label{app:derivation}

As it was mentioned in Sec. \ref{sec:preliminaries}, the mathematical methods we have brought here are strongly similar to those which have been discussed rigorously in Ref. \cite{Thompson2017}. The Raychaudhuri equation gives the evolution of the fractional rate of change of the transverse subspace, i.e. the congruence expansion. In the language we used in Sec. \ref{sec:preliminaries}, and using Eq. (\ref{four}) for the expansion, this evolution is calculated as
\begin{eqnarray}\label{app:I}
\dot\Theta = \dot{\bar B}{^\mu}_\mu &=& \frac{D}{\ed \tau}\left(h{_\rho}^\mu \ell{^\rho}_{;\beta}h{_\mu}^\beta\right)\nonumber\\
&=& h{_\rho}^\mu \frac{D}{\ed\tau}\left(\ell{^\rho}_{;\beta}\right)h{_\mu}^\beta
+\frac{D}{\ed\tau}\left(h{_\rho}^\mu h{_\mu}^\beta\right)\ell{^\rho}_{;\beta}
=h{_\rho}^\mu\left(\ell^\tau\ell{^\rho}_{;\beta\tau}\right)h{_\mu}^\beta 
+ \ell^\tau\left(h{_\rho}^\beta h{_\mu}^\beta\right)_{;\tau}\ell{^\rho}_{;\beta}\nonumber\\
&=& h{_\rho}^\mu \left(\ell^\tau \ell{^\rho}_{;\tau\beta} - R{^\rho}_{\lambda\beta\tau}\ell^\lambda\ell^\tau\right)h{_\mu}^\beta
+\ell^\tau\left(h{_\rho}^\mu h{_\mu}^\beta\right)_{;\tau}\ell{^\rho}_{;\beta}\nonumber\\
&=& h{_\rho}^\mu\left(a{^\rho}_{;\beta} - \ell{^\rho}_{;\tau}\ell{^\tau}_{;\beta}-R{^\rho}_{\lambda\beta\tau}\ell^\lambda\ell^\tau\right)h{_\mu}^\beta
+\ell^\tau\left(h{_\rho}^\beta\right)_{;\tau}\ell{^\rho}_{;\beta}\nonumber\\
&=&h{_\rho}^\beta a{^\rho}_{;\beta} - h{_\rho}^\mu\ell{^\rho}_{;\tau}\ell{^\tau}_{;\beta}h{_\mu}^\beta - h{_\rho}^\beta R{^\rho}_{\lambda\beta\tau}\ell^\lambda\ell^\tau + \dot{h}{_\rho}^\beta \ell{^\rho}_{;\beta}.
\end{eqnarray}
Expanding the second term, we have
\begin{eqnarray}\label{app:II}
 h{_\rho}^\mu\ell{^\rho}_{;\tau}\ell{^\tau}_{;\beta}h{_\mu}^\beta
 &=& \left(h{_\rho}^\mu\ell{^\rho}_{;\tau}h{_\nu}^\tau\right)\left(h{_\sigma}^\nu\ell{^\sigma}_{;\beta}h{_\mu}^\beta
 \right)
 +h{_\rho}^\mu\ell{^\rho}_{;\tau}\left(\delta_\nu^\tau-h{_\nu}^\tau\right)
 \left(\delta_\sigma^
 \nu-h{_\sigma}^\nu\right)\ell{^\sigma}_{;\beta}h{_\mu}^\beta\nonumber\\
 &=&\bar B{^\mu}_\nu\bar B{^\nu}_\mu + h{_\rho}^\mu\ell{^\rho}_{;\tau}\left(\delta_\nu^\tau-h{_\nu}^\tau\right)\left(\delta_\sigma^\nu-h{_\sigma}^\nu\right)\ell{^\sigma}_{;\beta}h{_\mu}^\beta,
\end{eqnarray}
with $\bar B{^\mu}_\nu$ to be the transverse projection of $B{^\mu}_\nu$. Using the definition given in Eq.~(\ref{five}), the third term of Eq.~(\ref{app:I}) yields
\begin{eqnarray}\label{app:III}
h{_\rho}^\beta R{^\rho}_{\lambda\beta\tau}\ell^\lambda\ell^\tau &=& R_{\lambda\tau}\ell^\lambda\ell^\tau - (\bm n\cdot\bm\ell)^{-1}\left[n_\rho\ell^\beta R{^\rho}_{\lambda\beta\tau}\ell^\lambda\ell^\tau+\ell_\rho n^\beta R{^\rho}_{\lambda\beta\tau}\ell^\lambda\ell^\tau\right]\nonumber\\
&=& R_{\lambda\tau}\ell^\lambda\ell^\tau,
\end{eqnarray}
since the terms in the brackets vanish by the virtue of anti-symmetry of the Rimemann tensor. Applying the above results in Eq.~(\ref{app:I}) gives
\begin{eqnarray}\label{app:IV}
\dot\Theta &=& -\bar{B}{^\mu}_\nu\bar{B}{^\nu}_\mu - R_{\lambda\tau}\ell^\lambda\ell^\tau + a{^\rho}_{;\rho} - (\bm n\cdot\bm\ell)^{-1}\left[n_\rho a{^\rho}_{;\beta}\ell^\beta + \ell_\rho a{^\rho}_{;\beta}n^\beta\right]\nonumber\\
&&-h{_\rho}^\mu\ell{^\rho}_{;\tau}\left(\delta_\nu^\tau-h{_\nu}^\tau\right)\left(\delta_\sigma^\nu-h{_\sigma}^\nu\right)\ell{^\sigma}_{;\beta}h{_\mu}^\beta + \dot{h}{_\rho}^\beta\ell{^\rho}_{;\beta}.
\end{eqnarray}
Now consider the term $h{_\rho}^\mu\ell{^\rho}_{;\tau}\left(\delta_\nu^\tau-h{_\nu}^\tau\right)\left(\delta_\sigma^\nu-h{_\sigma}^\nu\right)\ell{^\sigma}_{;\beta}h{_\mu}^\beta$. This term is equivalent to
\begin{eqnarray}\label{app:V}
h{_\rho}^\mu \ell{^\rho}_{;\tau}\ell{^\nu}_{;\mu}\left(\delta_\nu^\tau - h{_\nu}^\tau\right)&=& \left[\delta_\rho^\mu - (\bm n\cdot\bm\ell)^{-1}\left(n_\rho\ell^\mu+\ell_\rho n^\mu\right)\right]
\left[(\bm n\cdot\bm\ell)^{-1}\left(n_\nu\ell^\tau + \ell_\nu n^\tau\right)
\right]\ell{^\nu}_{;\mu}\ell{^\rho}_{;\tau}\nonumber\\\nonumber\\
&=& \left\{
(\bm n\cdot\bm\ell)^{-1}\delta_\rho^\mu n_\nu\ell^\tau +(\bm n\cdot\bm\ell)^{-1}
\delta_\rho^\mu \ell_\nu n^\tau
-(\bm n\cdot\bm\ell)^{-2}\left[n_\rho\ell^\mu n_\nu \ell^\tau + n_\rho\ell^\mu\ell_\nu n^\tau
\right.\right.\nonumber\\
&&\left.\left.+
\ell_\rho n^\mu n_\nu \ell^\tau + \ell_\rho n^\mu\ell_\nu n^\tau
\right]\right\} \ell{^\nu}_{;\mu}\ell{^\rho}_{;\tau}\nonumber\\\nonumber\\
&=& (\bm n\cdot\bm\ell)^{-1} \ell{^\nu}_{;\rho} \ell{^\rho}_{;\tau}
n_\nu \ell^\tau + (\bm n\cdot\bm\ell)^{-1}\ell{^\nu}_{;\rho}\ell{^\rho}_{;\tau}\ell_\nu n^\tau\nonumber\\
&&-(\bm n\cdot\bm\ell)^{-2}\ell{^\nu}_{;\mu}\ell{^\rho}_{;\tau}n_\rho\ell^\mu n_\nu \ell^\tau\nonumber
-(\bm n\cdot\bm\ell)^{-2}\ell{^\nu}_{;\mu}\ell{^\rho}_{;\tau}n_\rho\ell^\mu\ell_\nu n^\tau\nonumber\\
&&-(\bm n\cdot\bm\ell)^{-2}\ell{^\nu}_{;\mu}\ell{^\rho}_{;\tau}\ell_\rho n^\mu n_\nu \ell^\tau
- (\bm n\cdot\bm\ell)^{-2} \ell{^\nu}_{;\mu}\ell{^\rho}_{;\tau}\ell_\rho n^\mu\ell_\nu n^\tau\nonumber\\\nonumber\\
&=& (\bm n\cdot\bm\ell)^{-1} n_\nu\ell{^\nu}_{;\rho}a^\rho - 
(\bm n\cdot\bm\ell)^{-2} n_\nu a^\nu n_\rho a^\rho,
\end{eqnarray}
in which in order to get to the last line, we used the fact that $\ell_{\rho}\ell{^\rho}_{;\mu} = \frac{1}{2}\left(\ell^\rho\ell_\rho\right)_{;\mu}=0$. On the other hand, plugging  
\begin{equation}\label{app:VI}
\dot{h}{_\rho}^\beta = -\left(\frac{D}{\ed\tau}(\bm n\cdot\bm\ell)^{-1}\right)\left(\ell_\rho n^\beta + n_\rho \ell^\beta\right)-(\bm n\cdot\bm\ell)^{-1}\left(a_\rho n^\beta + \ell_\rho b^\beta + b_\rho \ell^\beta + n_\rho a^\beta\right)
\end{equation}
into the last term of Eq.~(\ref{app:I}), yields
\begin{eqnarray}\label{app:VII}
\dot{h}{_\rho}^\beta \ell{^\rho}_{;\beta} &=& \left[
\dot{(\bm n\cdot\bm\ell)}(\bm n\cdot\bm\ell)^{-2}\left(\ell_\rho n^\beta + n_\rho\ell^\beta\right)-(\bm n\cdot\bm\ell)^{-1}\left(a_\rho n^\beta + \ell_\rho b^\beta+b_\rho\ell^\beta + n_\rho a^\beta\right)
\right]\ell{^\rho}_{;\beta}\nonumber\\
&=&  (\bm n\cdot\bm\ell)^{-2}\dot{(\bm n\cdot\bm\ell)} n_\rho a^\rho
-(\bm n\cdot\bm\ell)^{-1}\left(a_\rho a_n^\rho + b_\rho a^\rho + n_\rho \ell{^\rho}_{;\beta}a^\beta\right).
\end{eqnarray}
Inserting Eqs. (\ref{app:V}) and (\ref{app:VII}) into Eq. (\ref{app:IV}), we reach
\begin{eqnarray}\label{app:VIII}
\dot \Theta &=& -\bar{B}{^\mu}_\nu\bar{B}{^\nu}_\mu - R_{\lambda\tau}\ell^\lambda\ell^\tau + a{^\rho}_{;\rho} - (\bm n\cdot\bm\ell)^{-1}\left[
n_\rho a{^\rho}_{;\beta}\ell^\beta + \ell_\rho a{^\rho}_{;\beta}n^\beta
\right]\nonumber\\
&&-(\bm n\cdot\bm\ell)^{-1} n_\nu\ell{^\nu}_{;\rho}a^\rho + (\bm n\cdot\bm\ell)^{-2}n_\nu a^\nu n_\rho a^\rho+(\bm n\cdot\bm\ell)^{-2}\dot{(\bm n\cdot\bm\ell)}n_\rho a^\rho\nonumber\\
&&-(\bm n\cdot\bm\ell)^{-1}\left(a_\rho a_n^\rho + b_\rho a^\rho + n_\rho \ell{^\rho}_{;\beta} a^\beta\right)\nonumber\\\nonumber\\
&=& -\bar{B}{^\mu}_\nu\bar{B}{^\nu}_\mu- R_{\lambda\tau}\ell^\lambda\ell^\tau + a{^\rho}_{;\rho} - a{^\rho}_{;\beta}\left(\delta_\rho^\beta - h{_\rho}^\beta\right)\nonumber\\
&&-(\bm n\cdot\bm\ell)^{-1}\left[
2n_\nu\ell{^\nu}_{;\rho}a^\rho - (\bm a\cdot\bm a_n)-\dot{(\bm b\cdot\bm a )}
\right]+(\bm n\cdot\bm\ell)^{-2}\left[(\bm n\cdot\bm a)^2 + \dot{(\bm n\cdot \bm \ell)}(\bm n\cdot \bm a)\right],
\end{eqnarray}
as the generalized Raychaudhuri equation. Note that, applying the relations in Eqs. (\ref{eight}) and (\ref{nine}), one gets
\begin{eqnarray}\label{app:IX}
\bar{B}{^\mu}_\nu\bar{B}{^\nu}_\mu &=&
\frac{1}{2}\Theta^2 + \sigma{^\mu}_\nu\sigma{^\nu}_\mu + \omega{^\mu}_\nu\omega{^\nu}_\mu\nonumber\\
&=& \frac{1}{2}\Theta^2 + \sigma_{\mu\nu}\sigma^{\mu\nu} - \omega_{\mu\nu}\omega^{\mu\nu}.
\end{eqnarray}
Interpolation of the above quantity in Eq.~(\ref{app:VIII}) gets us to Eq.~(\ref{III}). To obtain the expansion relation in Eq.~(\ref{I}), let us expand the following relation:
\begin{eqnarray}\label{app:X}
\Theta &=& h{_\sigma}^\rho \ell{^\sigma}_{;\rho}\nonumber\\
&=& \left[\delta_\sigma^\rho - (\bm n\cdot\bm\ell)^{-1}\left(
\ell_\sigma n^\rho + n_\sigma\ell^\rho
\right)\right]\ell{^\sigma}_{;\rho}\nonumber\\
&=&\ell{^\rho}_{;\rho}-(\bm n\cdot\bm\ell)^{-1} n_\sigma\ell{^\sigma}_{;\rho}\ell^\rho = \ell{^\rho}_{;\rho}-(\bm n\cdot\bm\ell)^{-1}  (\bm n\cdot\bm a),
\end{eqnarray}
in which the $n_\sigma\ell{^\sigma}_{;\rho} \ell^\rho$ in the third line can be recast as
\begin{eqnarray}\label{app:XI}
n_\sigma\ell{^\sigma}_{;\rho} \ell^\rho &=& \left[
\left(
\bm n\cdot\bm\ell
\right)_{;\rho}
-\ell^\sigma n_{\sigma;\rho}
\right]\ell^\rho\nonumber\\
&=& \dot{\left(
\bm n\cdot\bm\ell
\right)} - (\bm\ell\cdot\bm b).
\end{eqnarray}
Therefore, for $\bm n\cdot\bm\ell=\mathrm{const.}$, one gains $\bm n\cdot \bm a = - \bm\ell\cdot\bm b$. Furthermore, the ingoing expansion can also be obtained by manipulating
\begin{eqnarray}\label{app:XII}
\Theta_n &=& h{_\sigma}^\rho n{^\sigma}_{;\rho}\nonumber\\
&=& \left[\delta_\sigma^\rho - (\bm n\cdot\bm\ell)^{-1}\left(
\ell_\sigma n^\rho + n_\sigma\ell^\rho
\right)\right]n{^\sigma}_{;\rho}\nonumber\\
&=&n{^\rho}_{;\rho}-(\bm n\cdot\bm\ell)^{-1} \ell_\sigma n{^\sigma}_{;\rho}n^\rho = n{^\rho}_{;\rho}-(\bm n\cdot\bm\ell)^{-1}  (\bm \ell\cdot\bm b_n),
\end{eqnarray}
in which the term $\ell_\sigma n{^\sigma}_{;\rho}n^\rho$ can be rewritten as
\begin{eqnarray}\label{app:XIII}
\ell_\sigma n{^\sigma}_{;\rho}n^\rho &=& \left[(\bm n\cdot\bm\ell)_{;\rho} - \ell_{\sigma;\rho}n^\sigma\right]n^\rho\nonumber\\
&=& (\bm n\cdot\bm\ell)_{;\rho} n^\rho - (\bm a_n\cdot\bm n),
\end{eqnarray}
implying that for normalized congruences, one gets $\bm\ell\cdot\bm b_n = -\bm n\cdot\bm a_n$. Another important issue, mentioned in Sec. \ref{sec:preliminaries}, was the reduction of the generalized Raychaudhuri equation for non-affinely parameterized congruences. For such congruences which are indeed normalized, one requires $a^\alpha = \kappa \ell^\alpha$. Imposing this in Eq.~(\ref{app:VIII}), together with Eq. (\ref{app:IX}), yields
\begin{eqnarray}\label{app:XIV}
\dot\Theta &=& -\frac{1}{2}\Theta^2 -  \sigma_{\mu\nu}\sigma^{\mu\nu} +\omega_{\mu\nu}\omega^{\mu\nu} - R_{\mu\nu} \ell^\mu\ell^\nu + \kappa\ell{^\rho}_{;\rho} - \kappa \ell{^\rho}_{;\beta}\left(\delta_\rho^\beta - h{_\rho}^\beta\right)\nonumber\\
&&-(\bm n\cdot\bm\ell)^{-1}\kappa \left[
2 (\bm n\cdot\bm a)
+ n_{\mu;\nu}\ell^\mu\ell^\nu
\right]
+\kappa^2.
\end{eqnarray}
Using Eq.~(\ref{V}), we have $\kappa\ell{^\rho}_{;\rho} = \kappa \Theta + \kappa^2$. Hence, Eq.~(\ref{XIV}) becomes
\begin{eqnarray}\label{app:XV}
\dot\Theta &=& -\frac{1}{2}\Theta^2 -  \sigma_{\mu\nu}\sigma^{\mu\nu} +\omega_{\mu\nu}\omega^{\mu\nu} - R_{\mu\nu} \ell^\mu\ell^\nu + \kappa\Theta + \kappa^2 - \kappa \left\{
\Theta + \kappa -\left[
\delta_\rho^\beta - (\bm n\cdot\bm\ell)^{-1}\left(
\ell_\rho n^\beta + n_\rho\ell^\beta
\right)
\right]\ell{^\rho}_{;\beta}
\right\}\nonumber\\
&&-(\bm n\cdot\bm\ell)^{-1}\left\{
2\kappa^2(\bm n\cdot\bm\ell) - \kappa
 n_\rho \ell^{\rho}_{;\beta}
\ell^\beta
\right\} + \kappa^2\nonumber\\\nonumber\\
&=&-\frac{1}{2}\Theta^2 -  \sigma_{\mu\nu}\sigma^{\mu\nu} +\omega_{\mu\nu}\omega^{\mu\nu} - R_{\mu\nu} \ell^\mu\ell^\nu + \kappa\Theta + \kappa^2\nonumber\\
&&-\kappa\left\{
\Theta + \kappa - \ell{^\rho}_{;\rho}+(\bm n\cdot\bm \ell)^{-1}\left(
\ell_\rho n^\beta + n_\rho \ell^\beta
\right)\ell{^\rho}_{;\beta}
\right\}\nonumber\\
&&-(\bm n\cdot\bm\ell)^{-1}\left[
2\kappa^2 (\bm n\cdot\bm\ell)-\kappa (\underbrace{\bm n\cdot\bm a}_{=\kappa (\bm n\cdot\bm \ell)})
\right] +\kappa^2\nonumber\\\nonumber\\
&=&  -\frac{1}{2}\Theta^2 -  \sigma_{\mu\nu}\sigma^{\mu\nu} +\omega_{\mu\nu}\omega^{\mu\nu} - R_{\mu\nu} \ell^\mu\ell^\nu + \kappa\Theta + \kappa^2 \nonumber\\
&&-\kappa\left[
(\bm n\cdot\bm \ell)^{-1}(\bm n\cdot\bm a)
\right]-(\bm n\cdot\bm \ell)^{-1}\left[\kappa^2(\bm n\cdot\bm \ell)\right] + \kappa^2\nonumber\\\nonumber\\
&=& -\frac{1}{2}\Theta^2 -  \sigma_{\mu\nu}\sigma^{\mu\nu} +\omega_{\mu\nu}\omega^{\mu\nu} - R_{\mu\nu} \ell^\mu\ell^\nu + \kappa\Theta. 
\end{eqnarray}

\section{Technical Note on Positivity and Negativity of the Components}\label{app:congruences}

It is desirable to distinguish the characteristics of $\bm\ell$ and $\bm n_{1,2}$. As it was pointed out before, for $\bm\ell$ to be future directed (more precisely, toward $+z$ and $+t$),  Eq. (\ref{XIII}) implies $-2 E H>0$ and $|2 E H|>|E|$. The above conditions require $H$ and $E$ to be of different signs and also $|H|>1/2$. However, the third component of $\bm\ell$ implies that $H$ must be positive and consequently, $E<0$. Therefore for $\bm\ell$ to be toward $+x$, we need the third component to be positive. Hence, we choose the $(-)$ sign of it. We introduced two cases of auxiliary vectors in Eqs. (\ref{XIV}). We inspect them in more details.

\begin{itemize}
\item{For $\bm n_1$ we consider the two conditions of $n_1^1>0$ (for $n_1^1$ to be along $+\bm v$) and $|n_1^1|>|n_1^0|$ (for $\bm n_1$ to be future-directed). The first inequality above is trivially satisfied. However, for the known conditions $E<0$ and $H>1/2$, the second inequality is impossible to be obviated. Therefore, $\bm n_1$ is not an associated future-directed congruence with $\bm\ell$ and we must disregard it. }

\item{Similar conditions have to be inspected for the second component of $\bm n_2$ and from Eq. (\ref{XIVb}) they indeed lead to the inequality
\begin{equation}\label{twenty}
2 \sqrt{E H n^0}+1>E (1-2 H) n^0,
\end{equation}
which since $H>1/2$, is equivalent to $\sqrt{2 E n^0}+1>n^0 E\times ( \text{something negative})$. Therefore, the above inequality also reaches a triviality and hence, $\bm n_2$ is future-directed. Now since this causal condition is respected, we turn to imposing a condition on $\bm n_2$, for it to be ingoing toward $-x$. This needs $n_2^2<0$, which according to Eq. (\ref{XIVb}), gives the inequality $\sqrt{H} n^0+\sqrt{\frac{n^0}{E}}>0$. As it was mentioned before, $n^0$ and $E$ must have the same signs (i.e. $n^0<0$ and we take it to be constant for the sake of convenience). Imposing the lower limit of $H$, the above relation reduces to $E n^0<2$. For this condition, we can regard $\bm n_2$ as a future-directed congruence, ingoing toward $-x$ direction.
}

\end{itemize}

%\bibliography{/data/Physics/Research/Papers/TransOptics/TransOptics,/data/Physics/Research/Papers/BlackHoles/AnalogModels/Analogs,/data/Physics/Research/BibFiles/Common,/data/Physics/Research/BibFiles/GR}

\end{document}